\documentclass[a4paper]{jpconf}

\usepackage{latexsym}
\usepackage{amsmath}
\usepackage{amssymb}
\usepackage{amsfonts}

\usepackage{color}

\usepackage{supertabular} 
\usepackage{placeins}
\usepackage{epsfig}
\usepackage{graphicx}

\usepackage{dsfont}
\usepackage{capt-of}

\usepackage{hyperref}

\begin{document}

\title{Determination of electric dipole transitions in heavy quarkonia using 
potential non-relativistic QCD}

\author{Jorge Segovia and Sebastian Steinbei\ss er}
\address{Physik-Department, Technische Universit\"at M\"unchen, \\ 
James-Franck-Str. 1, 85748 Garching, Germany}
\ead{jorge.segovia@tum.de, sebastian.steinbeisser@tum.de}

\begin{abstract}
The electric dipole transitions $\chi_{bJ}(1P)\to \gamma\Upsilon(1S)$ with 
$J=0,1,2$ and $h_{b}(1P)\to \gamma\eta_{b}(1S)$ are computed using the 
weak-coupling version of a low-energy effective field theory named potential 
non-relativistic QCD (pNRQCD). In order to improve convergence and thus give 
firm predictions for the studied reactions, the full static potential is 
incorporated into the leading order Hamiltonian; moreover, we must handle 
properly renormalon effects and re-summation of large logarithms. The precision 
we reach is $k_{\gamma}^{3}/(mv)^{2} \times \mathcal{O}(v^{2})$, where 
$k_{\gamma}$ is the photon energy, $m$ is the mass of the heavy quark and $v$ 
its velocity. Our analysis separates those relativistic contributions that 
account for the electromagnetic interaction terms in the pNRQCD Lagrangian which 
are $v^{2}$ suppressed and those that account for wave function corrections of 
relative order $v^{2}$. Among the last ones, corrections from $1/m$ and $1/m^2$ 
potentials are computed, but not those coming from higher Fock states since 
they demand non-perturbative input and are $\Lambda_{\text{QCD}}^{2}/(mv)^{2}$ 
or $\Lambda_{\text{QCD}}^{3}/(m^{3}v^{4})$ suppressed, at least, in the strict 
weak coupling regime. These proceedings are based on the forthcoming 
publication~\cite{steinbeisser}.
\end{abstract}

%%%%%%%%%%%%%%%%%%%%%%%%%%%%%%%%%%%%%%%%%%%%%%%%%%%%%%%%%%%%%%%%%%%%%%%%%%%%%%%
%%%%%%%%%%%%%%%%%%%%%%%%%%%%%%%%%%%%%%%%%%%%%%%%%%%%%%%%%%%%%%%%%%%%%%%%%%%%%%%
\vspace*{-0.60cm}

\section{Introduction}
\label{sec:introduction}

The electric dipole (E1) and magnetic dipole (M1) transitions between heavy 
quarkonia have been treated for a long time by means of potential models that 
use non-relativistic reductions of QCD-based quark--(anti-)quark interactions 
(see, for instance, Ref.~\cite{Segovia:2016xqb} for a recent application to the 
bottomonium system). However, a new large set of accurate experimental data 
related with electromagnetic reactions in the heavy quark sector is expected to 
be reported by B-factories (Belle@KEK), $\tau$-charm facilities (BES@IHEP) and 
even proton--(anti-)proton colliders (LHCb@CERN and PANDA@GSI); therefore, 
demanding for a systematic and model-independent analysis.

The aim of this manuscript is to compute the E1 transitions $\chi_{bJ}(1P)\to 
\gamma\Upsilon(1S)$ with $J=0,1,2$ and $h_{b}(1P)\to \gamma\eta_{b}(1S)$ using 
the effective field theory (EFT) called pNRQCD~\cite{Pineda:1997bj, 
Brambilla:1999xf} (see Refs.~\cite{Brambilla:2004jw, Pineda:2011dg} for 
reviews). This EFT takes full advantage of the hierarchy of scales that appear 
in the system:\footnote{The heavy quark mass is $m$, the relative momentum of 
the heavy quarks is $p$ and the bound state energy is denoted by $E$. The heavy 
quark velocity, $v$, is assumed to be $v\ll1$ which is reasonably fulfilled in 
bottomonium ($v^{2} \sim 0.1$).} $m \gg p \sim 1/r \sim mv  \gg E \sim mv^2\,,$ 
and makes a systematic connection between the underlying quantum field theory 
and non-relativistic quantum mechanics.

The specific details on the construction of pNRQCD depend on the relative size 
of the $mv$ scale with respect to $\Lambda_{\text{QCD}}$, for $m v\gg 
\Lambda_{\text{QCD}}$ we have the weak-coupling regime~\cite{Pineda:2011dg}, 
and for $mv \gtrsim \Lambda_{\text{QCD}}$ the strong-coupling 
regime~\cite{Brambilla:2000gk}. 

Nowadays, there seems to be a growing consensus that the weak-coupling regime 
works properly for many physical observables in the bottomonium sector. In 
order to reach this conclusion, one must (i) include the static potential in 
the leading order Hamiltonian, which produces a more convenient rearrangement 
of the perturbative series; (ii) treat adequately the renormalon effects, 
which improves the convergence of the series; and (iii) calculate the 
re-summation of large logarithms, which significantly diminish the 
factorization scale dependence of the observable.

The improvements mentioned above have been applied already to the determination 
of M1 transitions between low-lying heavy quarkonium states in  
Ref.~\cite{Pineda:2013lta}. Therein, good convergence properties and agreement 
between theory and experiment was obtained for the bottomonium ground state and 
for the $n=2$ excitation of the bottomonium system in both $S$- and $P$-waves. 
The motivation of the present paper is to repeat the study of 
Ref.~\cite{Pineda:2013lta} to the case of E1 transitions between bottomonia. In 
this case, contrary to M1 transitions, the computation of relativistic 
corrections is technically complicated: in addition to the effects 
given by higher order terms in the E1 transition operator, one has to calculate 
many corrections to the initial and final state wave function due to higher 
order potentials and higher order Fock states. This fact has hindered numerical 
computations of the E1 transitions between low-lying heavy quarkonium states 
within pNRQCD (see Refs.~\cite{Pietrulewicz:2013ct, Martinez:2016spe} for 
partial calculations) and this contribution aims to close the gap.

%%%%%%%%%%%%%%%%%%%%%%%%%%%%%%%%%%%%%%%%%%%%%%%%%%%%%%%%%%%%%%%%%%%%%%%%%%%%%%%
%%%%%%%%%%%%%%%%%%%%%%%%%%%%%%%%%%%%%%%%%%%%%%%%%%%%%%%%%%%%%%%%%%%%%%%%%%%%%%%
\vspace*{-0.20cm}

\section{Theoretical setup}
\label{sec:theory}

The formulae of the E1 transitions in the weak-coupling regime of pNRQCD have 
been presented in detail in Ref.~\cite{Brambilla:2012be}. Up to order 
$k_{\gamma}^{3}/m^{2}$, the expressions we use for the decay rates under study 
are
\begin{align}
\label{eq:FullDecayWidth}
&
\Gamma(n\,^3\!P_J \to n'\,^3\!S_1 + \gamma) = \Gamma_{E1}^{(0)}\, \Bigg\{ 1 + 
R^{S=1}(J) - \frac{k_\gamma}{6m} - \frac{k_\gamma^2}{60} \frac{I_5^{(0)}(n1 \to 
n'0)}{I_3^{(0)}(n1 \to n'0)} \nonumber \\
& 
\hspace*{4.80cm} + \left[ \frac{J(J+1)}{2} - 2 \right] \Bigg[ 
-\left(1+\kappa_Q^\text{em}\right) \frac{k_\gamma}{2m} \nonumber \\
&
\hspace*{4.80cm} + \frac{1}{m^2} (1 + 2\kappa_Q^\text{em}) \frac{I_2^{(1)}(n1 
\to n'0) + 2I_1^{(0)}(n1 \to n'0)}{I_3^{(0)}(n1 \to n'0)} \Bigg] \Bigg\} \,, 
\\[1ex]
&
\Gamma(n^1\!P_1 \to n'\,^1\!S_0 + \gamma) = \Gamma_{E1}^{(0)}\, \Bigg\{ 1 + 
R^{S=0} - \frac{k_\gamma}{6m} - \frac{k_\gamma^2}{60} \frac{I_5^{(0)}(n1 \to 
n'0)}{I_3^{(0)}(n1 \to n'0)} \Bigg\} \,,
\label{eq:FullDecayWidth2}
\end{align}
where $R^{S=1}(J)$ and $R^{S=0}$ include the initial and final state 
corrections due to higher order potentials and higher order Fock states (see 
below). The remaining corrections within the brackets are the result of taking 
into account additional electromagnetic interaction terms in the Lagrangian 
suppressed by ${\cal O}(v^2)$~\cite{Brambilla:2012be}. We have displayed in the 
formulae terms proportional to the anomalous magnetic moment, 
$\kappa_{Q}^{\text{em}}$. These terms are not considered in the numerical 
analysis because they are at least suppressed by $\alpha_{s}(m)v^2$ and thus 
go beyond our accuracy. The LO decay width, which scales as $\sim 
k_{\gamma}^{3}/(mv)^2$, is
\begin{equation}
\Gamma_\text{E1}^{(0)} = \frac{4}{9}\, \alpha_\text{em}\, e_Q^2\, k_\gamma^3 
\left[I_3^{(0)}(n1 \to n'0) \right]^{2} \,,
\label{eq:Gamma0}
\end{equation}
with $\alpha_\text{em}$ the electromagnetic fine structure constant, $e_Q$ the 
charge of the heavy quarks in units of the electron charge, $k_\gamma$ the 
photon energy, and the function
\begin{equation}
I_N^{(k)}(n\ell \to n'\ell') = \int_0^\infty dr\, r^{N} R_{n'\ell'}^{\ast}(r) 
\left[ \frac{d^k}{d r^k} R_{n\ell}(r) \right]
\end{equation}
is a matrix element that involves the radial wave functions of the initial and 
final states. We assume that these states are solutions of the Schr\"odinger 
equation
\begin{equation}
\label{eq:SchEqu} 
H^{(0)} \psi_{n\l m}^{(0)}(\vec{r}\,) = E_n^{(0)} \psi_{n\l m}^{(0)}(\vec{r}\,) 
\,,
\end{equation}
with the leading order Hamiltonian given by
\begin{equation}
H^{(0)} = -\frac{\vec{\nabla}^2}{2m_r} + V_{s}^{(N)}(r) \,,
\end{equation}
and the static potential approximated by a polynomial of order $N+1$ in powers 
of $\alpha_{s}$
\begin{equation}
V_{s}^{(N)}(r) = -\frac{C_{F}\alpha_s(\nu)}{r} \left[ 1 + \sum\limits_{k=1}^{N} 
\left(\frac{\alpha_s}{4\pi}\right)^k a_k(\nu,r) \right] \,.
\label{eq:StatPot2}
\end{equation}
In principle, one would like to take $N$ as large as possible; in practice, we 
truncate the perturbative series at $N=3$, \textit{i.e.} at 
$\mathcal{O}(\alpha_s^4)$, including also the leading ultra-soft corrections.

Due to higher order potentials and the presence of ultra-soft gluons that lead 
to singlet-to-octet transitions, the state in Eq.~(\ref{eq:SchEqu}) is not an 
eigenstate of the complete Hamiltonian. Therefore, one has to consider 
corrections to the wave function which contribute to the decay rate at the 
required order of precision. The first kind of corrections to the wave function 
are those which arise from relativistic corrections to the potential and to the 
leading order kinetic operator. They can be organized as an expansion in the 
inverse of the heavy quark mass, $m$. At the order we are interested in, such 
expansion covers all the $1/m$ and $1/m^2$ potentials and, at order $1/m^{3}$, 
the first relativistic correction to the kinetic 
energy~\cite{Brambilla:2012be}. The second kind of corrections to the wave 
function come from diagrams in which a singlet state is coupled to an octet 
state due to the emission and re-absorption of an ultra-soft gluon. We do not 
consider these contributions herein because in the (strict) weak-coupling 
regime, $E\sim mv^{2} \gg \Lambda_{\text{QCD}}$, they are 
negligible~\cite{Brambilla:2012be}.

%%%%%%%%%%%%%%%%%%%%%%%%%%%%%%%%%%%%%%%%%%%%%%%%%%%%%%%%%%%%%%%%%%%%%%%%%%%%%%%
%%%%%%%%%%%%%%%%%%%%%%%%%%%%%%%%%%%%%%%%%%%%%%%%%%%%%%%%%%%%%%%%%%%%%%%%%%%%%%%
\vspace*{-0.20cm}

\section{Results}
\label{sec:results}

\begin{figure*}[!t]
\centering
\includegraphics[clip,width=0.44\textwidth,height=0.25\textheight]
{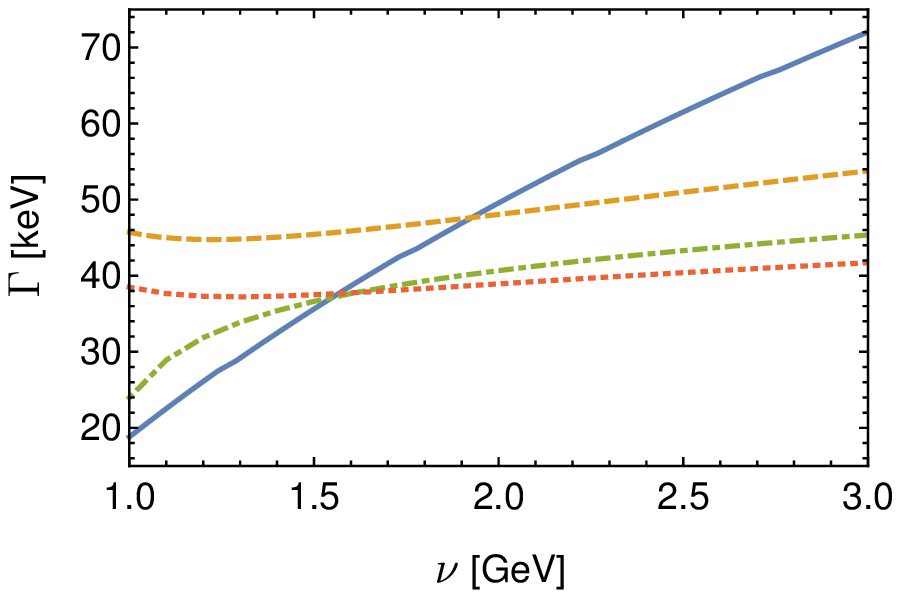}
\put(-160,150){\small (a)}
\hspace*{0.50cm}
\includegraphics[clip,width=0.44\textwidth,height=0.2575\textheight]
{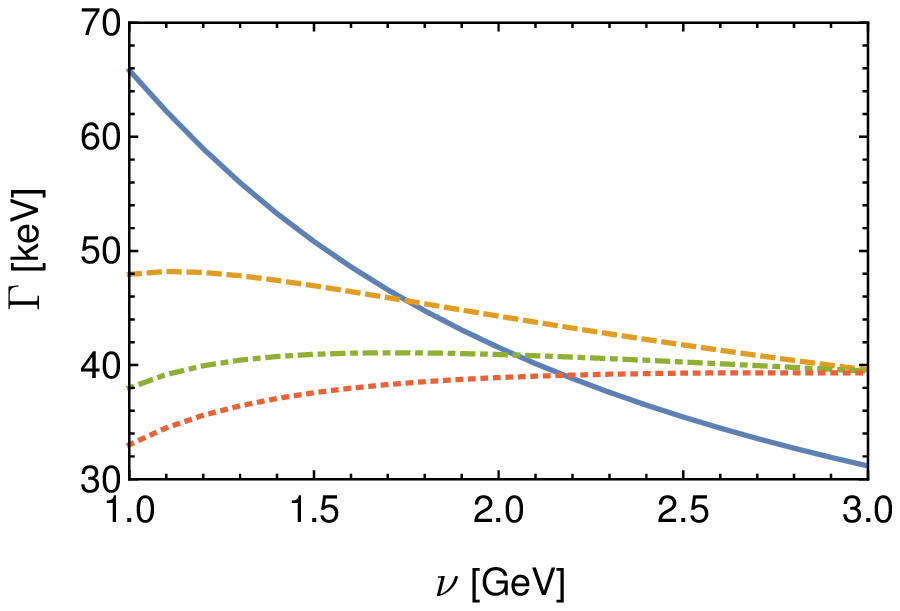}
\put(-160,150){\small (b)}
\caption{\label{fig:Renormalon+RGI} Leading order decay rate, 
$\Gamma_{E1}^{(0)}$, of the electric dipole transition $\chi_{b1}(1P)\to 
\gamma\Upsilon(1S)$. The renormalon cancellation is achieved order by order in 
the static potential which is exactly included when solving the Schr\"odinger 
equation. The panel (a) shows the case in which the renormalization group 
improved (RGI) version of the static potential is not used, whereas the panel 
(b) shows the results when the RGI potential is considered. The different 
curves in both panels represent the result coming from taking into account the 
Coulomb-like (solid blue), NLO (dashed orange), NNLO (dot-dashed green) and 
NNNLO (dotted red) terms in the static potential.}
\end{figure*}

The panel (a) of Fig.~\ref{fig:Renormalon+RGI} shows the leading order 
decay rate, $\Gamma_{E1}^{(0)}$, for the $\chi_{b1}(1P)\to \gamma\Upsilon(1S)$ 
transition computed when the static potential is included order by order in the 
Schr\"odinger equation. If one solves the Schr\"odinger equation numerically 
with only the Coulomb-like term of the static potential, one gets a leading 
order (LO) decay rate (solid blue curve) that depends strongly on the 
factorization scale (see Ref.~\cite{Steinbeisser:2017hke} for details based on 
an analytical study). However, the $\nu$-scale dependence becomes mild as 
next-to-leading order (NLO) (dashed orange curve), NNLO (dot-dashed green curve) 
and NNNLO (dotted red curve) radiative corrections to the static potential are 
included in the Schr\"odinger equation.

The panel (b) of Fig.~\ref{fig:Renormalon+RGI} shows the induced effect on 
the decay width once the correct log-arithmetically modulated short distance 
behavior (RGI) of the static potential is included properly. Now, the 
improvement in the convergence properties of the perturbative series is seen at 
$\nu\lesssim1.7\,{\rm GeV}$. This is crucial in order to give final results for 
the decay rates since a low value of $\nu$, compatible with the typical 
momentum scale of the system under study, has to be chosen 
($\nu=1.25\,\text{GeV}$). It is fair to mention that, in this final scheme, the 
dependence on the factorization scale for the NNNLO result seems to increase 
slightly.

\begin{figure}[!t]
% \centering
%
% \includegraphics[clip,width=0.44\textwidth,height=0.25\textheight]
% {G_Final_cb0_num.eps}
% \put(-90,43){\small $\chi_{b0}(1P)\to \gamma\Upsilon(1S)$}
%
\includegraphics[clip,width=0.44\textwidth,height=0.25\textheight]
{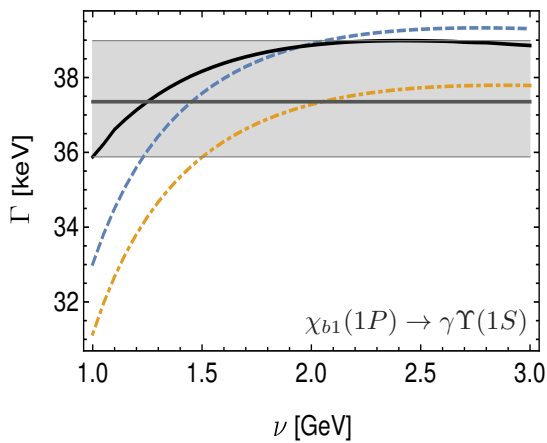}
\put(-90,43){\small $\chi_{b1}(1P)\to \gamma\Upsilon(1S)$}
\hspace*{0.60cm}
%
% \includegraphics[clip,width=0.44\textwidth,height=0.25\textheight]
% {G_Final_cb2_num.eps}
% \put(-90,43){\small $\chi_{b2}(1P)\to \gamma\Upsilon(1S)$}
%
% \includegraphics[clip,width=0.44\textwidth,height=0.25\textheight]
% {G_Final_hb_num.eps}
% \put(-87,43){\small $h_b(1P)\to \gamma\eta_b(1S)$}
%
\begin{minipage}[b]{18pc}
\caption{\label{fig:GfinalChib1} Decay width for the electric 
dipole transition $\chi_{b1}(1P)\to \gamma\Upsilon(1S)$. The dashed blue curve 
is the leading order decay rate, the dot-dashed orange curve is including the 
relativistic contributions stemming from higher order electromagnetic 
operators (Eq.~\eqref{eq:FullDecayWidth} with $R^{S=1}(J)=0$) and 
the solid black curve is the final result including also the relativistic 
corrections to the wave function of the initial and final states 
(Eq.~\eqref{eq:FullDecayWidth} with $R^{S=1}(J)\neq0$). We take our final value 
at $\nu = 1.25\,\text{GeV}$ and the gray band indicates the associated 
uncertainty.}
\vspace*{0.20cm}
\end{minipage}
\end{figure}

The final decay width for the $\chi_{b1}(1P)\to \gamma\Upsilon(1S)$ process is 
shown in Fig.~\ref{fig:GfinalChib1}. The leading order non-relativistic decay 
rate is the dashed blue curve, the dot-dashed curve is taking into account the 
relativistic contributions stemming from higher order electromagnetic operators 
(Eq.~\eqref{eq:FullDecayWidth} with $R^{S=1}(J)=0$) and the solid black one is 
including also the relativistic corrections to the wave function of the initial 
and final states (Eq.~\eqref{eq:FullDecayWidth} with $R^{S=1}(J)\neq0$). 

One sees in Fig.~\ref{fig:GfinalChib1} that the leading order decay width 
depends weakly on the factorization scale, it varies from $\Gamma\sim33\,\text{
keV}$ at $\nu=1\,\text{GeV}$ to $\Gamma\sim39\,\text{keV}$ at 
$\nu=3\,\text{GeV}$. This feature is translated to the case in which higher 
order electromagnetic operators are taken into account and also to the case in 
which wave function relativistic corrections are included. In fact, somewhat 
surprisingly, the $\nu$-dependence of our final result, $\sim3\,\text{keV}$, is 
weaker than that of the leading order or even the one including higher order 
electromagnetic operators. A variation of $\sim3\,\text{keV}$ over a total 
value of around $\sim37\,\text{keV}$ represents a relative error of $\sim8\%$ 
in our determination of the decay rate, being the biggest source of uncertainty.

It is clearly seen in Fig.~\ref{fig:GfinalChib1} that both types of 
relativistic contributions are under control: they behave smoothly with respect 
the renormalization scale and produce corrections to the LO decay width which 
are relatively small. Another interesting feature of Fig.~\ref{fig:GfinalChib1} 
is that the relativistic corrections induced by higher order electromagnetic 
operators tend to diminish the LO decay rate whereas the effect of the 
corrected initial and final wave functions is to increase it.

We have performed the same analysis than above to the electric dipole 
transitions $\chi_{b0}(1P)\to \gamma\Upsilon(1S)$, $\chi_{b2}(1P)\to 
\gamma\Upsilon(1S)$ and $h_{b}(1P)\to \gamma\Upsilon(1S)$. Similar conclusions 
than the ones already mentioned apply to these cases. Our final values for the 
electric dipole transitions $\chi_{bJ}(1P)\to \gamma\Upsilon(1S)$ with 
$J=0,1,2$ and $h_{b}(1P)\to \gamma\eta_{b}(1S)$ read
\begin{align}
\Gamma(\chi_{b0}(1P)\to \gamma\Upsilon(1S)) &= 28^{+2}_{-1}~\text{keV} \,, \\
\Gamma(\chi_{b1}(1P)\to \gamma\Upsilon(1S)) &= 37^{+2}_{-1}~\text{keV} \,, \\
\Gamma(\chi_{b2}(1P)\to \gamma\Upsilon(1S)) &= 45^{+1}_{-1}~\text{keV} \,, \\
\Gamma(h_b(1P)\to \gamma\eta_b(1S)) &= 63^{+1}_{-1}~\text{keV} \,.
\end{align}
Because of the very mild dependence of our results on the renormalization scale 
$\nu$, the associated uncertainties are very small.

The Particle Data Group (PDG)~\cite{Olive:2016xmw} only reports the branching 
fractions of the E1 transitions studied herein. Since the total decay widths of 
the $\chi_{bJ}$ (with $J=0,1,2$) and $h_{b}(1P)$ mesons are not known, we cannot 
compare our theoretical results with experimental data. Nevertheless, we can use 
the branching fractions given by the PDG and our results for the decay rates of 
the electric dipole transitions to predict their total decay widths. The results 
are
\begin{align}
\Gamma(\chi_{b0}(1P)) &= 1.6^{+0.3}_{-0.3}~\text{MeV} \,, \hspace*{1.85cm}
\Gamma(\chi_{b1}(1P))  = 110^{+9}_{-8}~\text{keV} \,, \\
\Gamma(\chi_{b2}(1P)) &= 234^{+15}_{-16}~\text{keV} \,, \hspace*{2.00cm}
\Gamma(h_b(1P))        = 121^{+14}_{-12}~\text{keV} \,.
\end{align}
These numbers could be of special interest for future experimental 
determinations. For instance, the Belle collaboration has recently reported an 
upper limit at 90\% confidence level on the total decay width of the 
$\chi_{b0}(1P)$~\cite{Abdesselam:2016xbr}: $\Gamma(\chi_{b0}(1P)) < 
2.4\,\text{MeV} $, which is well within the uncertainty band of our prediction.

%%%%%%%%%%%%%%%%%%%%%%%%%%%%%%%%%%%%%%%%%%%%%%%%%%%%%%%%%%%%%%%%%%%%%%%%%%%%%%%
%%%%%%%%%%%%%%%%%%%%%%%%%%%%%%%%%%%%%%%%%%%%%%%%%%%%%%%%%%%%%%%%%%%%%%%%%%%%%%%
\vspace*{-0.20cm}

\section{Summary}
\label{sec:summary}

We have performed the first numerical analysis of the electric dipole 
transitions $\chi_{bJ}(1P)\to \gamma\Upsilon(1S)$ with $J=0,1,2$ and 
$h_{b}(1P)\to \gamma\eta_{b}(1S)$ within the weak coupling version of a 
low-energy effective field theory called potential non-relativistic QCD.

%%%%%%%%%%%%%%%%%%%%%%%%%%%%%%%%%%%%%%%%%%%%%%%%%%%%%%%%%%%%%%%%%%%%%%%%%%%%%%%
%%%%%%%%%%%%%%%%%%%%%%%%%%%%%%%%%%%%%%%%%%%%%%%%%%%%%%%%%%%%%%%%%%%%%%%%%%%%%%%
\vspace*{-0.20cm}

\ack

This work has been supported by the DFG and the NSFC through funds provided to 
the Sino-German CRC 110 ``Symmetries and the Emergence of Structure in QCD'', 
and by the DFG cluster of excellence ``Origin and structure of the universe'' 
(www.universe-cluster.de). J.S. acknowledges the financial support from the 
Alexander von Humboldt Foundation.

%%%%%%%%%%%%%%%%%%%%%%%%%%%%%%%%%%%%%%%%%%%%%%%%%%%%%%%%%%%%%%%%%%%%%%%%%%%%%%%
%%%%%%%%%%%%%%%%%%%%%%%%%%%%%%%%%%%%%%%%%%%%%%%%%%%%%%%%%%%%%%%%%%%%%%%%%%%%%%%
\vspace*{-0.20cm}

\section*{References}

% Including bibliography through bibtex

\bibliographystyle{iopart-num}
\bibliography{proceedings_FAIRNESS2017_JorgeSegovia}

\end{document}